\documentclass[12pt]{article}
\pdfoutput=1

\usepackage{jheppub}
\usepackage{mathtools}
\usepackage{ulem}

\usepackage{etoolbox}
\usepackage{cancel}

\def\be{\begin{equation}}
\def\ee{\end{equation}}
\def\ba#1\ea{\begin{align}#1\end{align}}
\def\bg#1\eg{\begin{gather}#1\end{gather}}
\def\bm#1\em{\begin{multline}#1\end{multline}}
\def\bmd#1\emd{\begin{multlined}#1\end{multlined}}

\def\a{\alpha}

\def\d{\delta}

\def\h{\eta}

\def\ll{\lambda}

\def\m{\mu}

\def\p{\phi}

\def\q{\theta}

\def\t{\tau}

\def\y{\psi}

\def\la{\label}
\def\ci{\cite}
\def\re{\ref}
\def\er{\eqref}
\def\se{\section}
\def\sse{\subsection}

\def\fr{\frac}

\def\pa{\partial}
\def\td{\tilde}

\def\eq{\equiv}

\def\qu{\quad}
\def\qqu{\qquad}
\def\lt{\left}
\def\rt{\right}
\def\({\left(}
\def\){\right)}
\def\[{\left[}
\def\]{\right]}

\def\mo{{\mathcal O}}

\def\diff{{\rm diff}}
\def\on{{\rm on}}
\def\bdy{{\rm bdy}}

\newtoggle{editing}
\togglefalse{editing}

\iftoggle{editing}{%
	\usepackage[draft]{showlabels}
	\newcommand{\JS}[1]{{\textbf{\textcolor{red}{#1}}}}
	
	\newcommand{\JSout}[1]{{\textbf{\textcolor{red}{\sout{#1}}}}}
	\newcommand{\JSeqout}[1]{{\mathbf{\textcolor{red}{\cancel{#1}}}}}

	\newcommand{\BC}[1]{{\textbf{\textcolor{blue}{#1}}}}
	\newcommand{\XD}[1]{{\textbf{\textcolor{magenta}{#1}}}}

}{%
	\newcommand{\JS}[1]{#1}
	
	\newcommand{\JSout}[1]{}
	\newcommand{\JSeqout}[1]{}
	
	\newcommand{\BC}[1]{#1}
	\newcommand{\XD}[1]{#1}

}

\begin{document}

\subheader{SU-ITP-14/16 \\ SLAC-PUB-15984}
\title{Holographic Reconstruction of General Bulk Surfaces}
\author[a]{Bart{\l}omiej Czech,} 
\author[a]{Xi Dong,}
\author[b]{and James Sully}
\affiliation[a]{Stanford Institute for Theoretical Physics, Department of Physics, Stanford University, Stanford, CA 94305, USA}
\affiliation[b]{Theory Group, SLAC National Accelerator Laboratory, Stanford University, Menlo Park, CA 94025, USA}
\emailAdd{czech@stanford.edu}
\emailAdd{xidong@stanford.edu}
\emailAdd{jsully@slac.stanford.edu}

\abstract{{We propose a reconstruction of general bulk surfaces in any dimension in terms of {the} differential entropy in the boundary field theory. In particular, we extend the proof of {Headrick} et al. to calculate the area of a general class of surfaces, which have a 1-parameter foliation over a closed manifold. The area can be written in terms of {extremal} surfaces whose boundaries lie on ring-like regions in the field theory. We discuss when this construction has a description in terms of spatial entanglement entropy and suggest lessons for a more complete and covariant approach.}}
\maketitle

\se{Introduction}
Nearly two decades after the formulation of holographic duality \cite{Maldacena:1997re}, the question of how a gravitational spacetime is assembled from field theoretic degrees of freedom remains mysterious. The most intriguing aspect of this problem is the radial direction, which is not present in the field theory and, therefore, must have an emergent character. Starting from different initial points, one can derive a number of qualitative conclusions about the radial direction, many of which are contradictory or counter-intuitive. For example, the UV/IR relation \cite{Susskind:1998dq} suggests that traveling along the radial direction toward the center of a holographic spacetime is dual to following the field theory RG flow toward the deep infrared \cite{Akhmedov:1998vf, Balasubramanian:1999jd, deBoer:1999xf}. On the other hand, low energy effective field theory in the bulk, which assumes that radially separated regions of the spacetime contain independent degrees of freedom, implies that a spacetime dual to an excited pure state ends abruptly at a radial scale, which is the horizon of the corresponding black hole \cite{Braunstein:2009my, Almheiri:2012rt}. To shed light on this problem -- to understand how space emerges and what happens when one jumps into a black hole -- it is necessary to study the radial direction in a quantitative way.

The present paper is part of an effort to study this problem using the relation between spacetime and quantum entanglement in the boundary theory. The cornerstone of this relation is the Ryu-Takayanagi proposal \cite{Ryu:2006bv}, which posits that the entanglement entropy of a spatial region $\mathcal{R}$ in the field theory equals the area of a bulk minimal surface that asymptotes to the boundary of $\mathcal{R}$ in Planck units. This proposal has profound consequences. A simple argument due to Van Raamsdonk \cite{VanRaamsdonk:2010pw} (see also \cite{Swingle:2009bg, Swingle:2012wq}) shows that quantum entanglement in the field theory is a necessary condition for the connectedness of spacetime: when entanglement between complementary field theory regions is tuned to zero, the spacetime appears to pinch off into disconnected components. Adopting this viewpoint, Bianchi and Myers \cite{Bianchi:2012ev} conjectured that the relation
\begin{equation}
\frac{\rm Area}{4G} = S_{\rm ent}
\end{equation}
extends to arbitrary surfaces and serves to {define} the spacetime from field theoretic degrees of freedom. In three-dimensional anti-de Sitter space (AdS$_3$), a modified version of this conjecture was recently made quantitative in \cite{Balasubramanian:2013lsa}. Starting from a family $I_j$ of boundary intervals with corresponding Ryu-Takayanagi minimal surfaces (geodesics) $m_j$, the circumference of the spatial region lying outside all the curves $m_j$ is given by a novel boundary quantity dubbed `differential entropy':
\begin{equation}
\frac{\rm circumference}{4G} = \sum_j \big( S_{\rm ent}(I_j) - S_{\rm ent}(I_j \cap I_{j+1})\big) \, .
\label{oldformula}
\end{equation}
While the differential entropy is assembled from entanglement, each differential length element in formula~(\ref{oldformula}) arises from entanglement of a distinct region $I_j$ of the boundary theory. The union of these regions covers a Cauchy slice of the field theory. This suggests that radially separated regions of the bulk spacetime only appear to contain independent degrees of freedom in the low energy approximation, while in fact they are microscopically built up of the same fundamental ingredients.

This lesson is not particular to AdS$_3$. As shown in \cite{Myers:2014jia}, a generalization of formula~(\ref{oldformula}) characterizes codimension-2 bulk surfaces in non-AdS holographic backgrounds, in other theories of gravity including Lovelock theories, and in higher dimensional setups in which the bulk surfaces have planar symmetry. This wealth of examples \BC{was} explained in \cite{Headrick:progress} (see also \cite{Myers:joefest}), which \BC{gave} a robust, background-independent proof of the generalized version of eq.~(\ref{oldformula}). In the present paper, we use this proof to extend the scope of the differential entropy formula to those general surfaces in more than 3 bulk dimensions, which {admit a}
1-parameter foliation over a closed manifold. Part of our motivation is to identify insights that may guide us toward a covariant version of differential entropy, which does not rely on a special foliation of the bulk surface. 

In Sec.~\ref{proposal}, we write down our generalization of differential entropy (eq.~\ref{sdiff}) and specify the class of bulk surfaces whose areas it computes. The surfaces must admit {a 1-parameter} foliation over a closed manifold, so that the technique of the proof of \cite{Headrick:progress} applies. We review this proof and extend it to our case of interest in Sec.~\ref{proof}. In Sec.~\ref{examples} we present an example, {which involves a} numerical computation. {For many bulk surfaces formula~(\ref{sdiff}) involves not minimal but merely extremal surfaces, whose areas do not compute boundary \JS{spatial} entanglement entropies. We characterize when this happens in Sec.~\ref{extr}.} Our results contain hints, which will likely be useful for writing down a fully covariant formula for the differential entropy. We discuss these hints in Sec.~\ref{discussion}.  For completeness, in Appendix~\re{app} we describe the numerical methods used in the example.

\se{Proposal}
\label{proposal}

\subsection{Setup}
{We work in a $(d+1)$-dimensional bulk geometry dual to a $d$-dimensional field theory on the boundary. Consider a $(d-1)$-dimensional surface in the bulk. We assume that the surface is closed, spacelike, and sufficiently differentiable. Our goal is to compute the generalized area of this surface using boundary ingredients. By generalized area we mean the integral over the bulk surface of the same density, which -- if integrated over a minimal surface -- would have given a boundary entanglement entropy according to the Ryu-Takayanagi proposal  \cite{Ryu:2006bv}. In the case of Einstein gravity this density is a constant and the generalized area is ordinary surface area.}

{In principle, the boundary ingredients we use are entanglement entropies of spatial regions in the field theory. These correspond to generalized areas of minimal surfaces in the bulk. However, we also allow extremal rather than strictly minimal surfaces, even though their boundary interpretation is not known. We shall not distinguish minimal surfaces from extremal surfaces until Sec.~\ref{extr}, where we characterize when the latter make an appearance. We comment on a possible boundary interpretation of extremal but not minimal surfaces in Sec.~\ref{extrdisc}, but refer the reader to \cite{Czech:progress} for a more thorough discussion.}

\subsection{Strategy}
\label{strategy}
The main idea is to foliate the bulk surface by codimension-1 ``loops.''
Here and below we use ``surfaces'' to refer to $(d-1)$-dimensional manifolds and ``loops'' to refer to $(d-2)$-dimensional ones, even though they are actual loops only in $d=3$.  In Poincar\'e coordinates they could also look like infinite lines.  Note that these lines are not required to be straight, but if they are, we are in the special case considered in \ci{Myers:2014jia}.

This {codimension-1} foliation allows us to construct coordinates $(\ll,\h^a)$ on the surface such that each loop $K(\ll)$ is specified by a constant $\ll$.  Here $a$ runs from $1$ to $d-2$.  Our goal is to reconstruct
\be\la{garea}
A= \int d\ll \, d^{d-2}\h \, L(x^\m,\pa_\ll x^\m,\pa_{\h^\a} x^\m)\,,
\ee
where $x^\m$ denotes all bulk coordinates and $L$ is the generalized area element that depends only on $x^\m$ and their first derivatives.  The holographic entanglement entropy is given by minimizing the integral of $L$ subject to the usual {homology} constraint. {We call $L$ the \textit{generalized} area element because we allow both Einstein gravity and any theory of higher derivative gravity with 
\JS{an action principle} for the holographic entanglement entropy \cite{Hung:2011xb, Dong:2013qoa}.}

As a next step, construct for each $\ll$  an extremal surface $M(\ll)$ that is tangent to the original surface precisely at the loop $K(\ll)$. 
In other words, we need to extend the loop $K(\ll)$ in one additional dimension. 
We shall use $s$ to denote a coordinate in this new direction and for simplicity reuse the $\h^a$ coordinates as the other coordinates on the extremal surface $M(\ll)$. Finding an extremal surface requires solving an elliptic partial differential equation; demanding that $M(\ll)$ meet $K(\ll)$ and be tangent to the given bulk surface sets a Cauchy boundary condition. In its most general form, this problem is not guaranteed to have a solution and if it does, the solution may not depend continuously on the Cauchy data.\footnote{We thank the reviewer for clarifying this point to us.} In what follows we assume that the solution $M(\ll)$ exists. This assumption holds for a variety of shapes in the bulk; we discuss a representative series of examples in Sec.~\ref{examples}. Discontinuous dependence of solutions on the Cauchy data adds additional boundary terms to the differential entropy. Our example in Sec.~\ref{spheres} illustrates this complication while Sec.~\ref{secphases} discusses an example ``phase structure'' of discontinuously varying extremal surfaces.


Since each loop $K(\ll)$ is closed by construction, the {extremal} surface $M(\ll)$ generically encloses a ring-shaped region $R(\ll)$ on the asymptotic boundary. {(In general, the topology of the ``ring'' $R(\ll)$ depends on the topology of the ``loop'' $K(\ll)$.)
For our purposes, it is important that} the loop $K(\ll)$ separates the minimal surface $M(\ll)$ into two regions -- left $M_L(\ll)$ and right $M_R(\ll)$ -- and that they intersect the asymptotic boundary at loops $B_L(\ll)$ and $B_R(\ll)$, respectively.  Let us write the field theory region between any two loops $B_L$ and $B_R$ on the asymptotic boundary as $[B_L,B_R]$.  Then the ring-shaped region $R(\ll)$ mentioned earlier can be written as $[B_L(\ll),B_R(\ll)]$.  See Fig.~\re{figillus} for an illustration of this.

\begin{figure}[t]
\centering
\raisebox{1.2cm}{\includegraphics[width=0.35\textwidth]{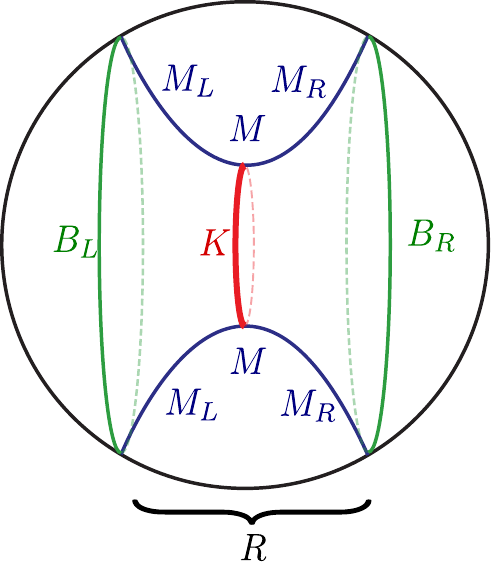}}
\includegraphics[width=0.54\textwidth]{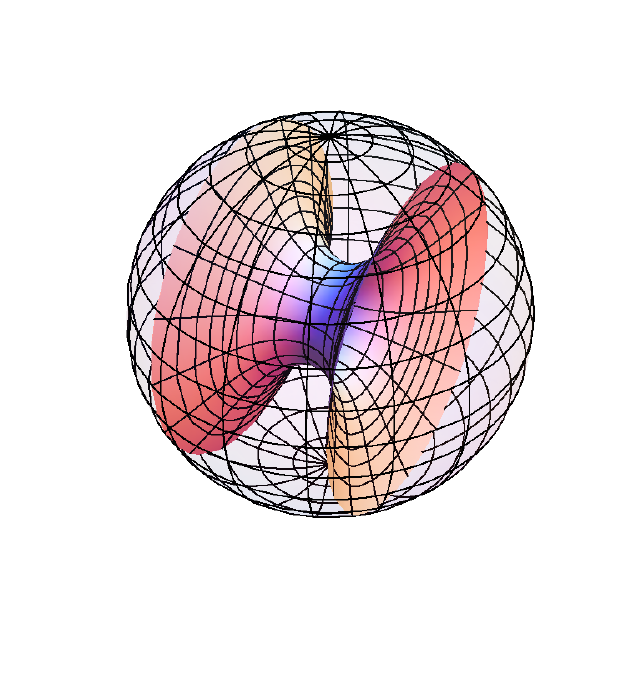}
\vspace{-1cm}
\caption{\label{figillus}A minimal surface $M$ tangent to the original surface (not drawn) at a loop $K$.
{The choice of which side is left/right is fixed by demanding that the coordinate $\ll$ increase from left to right.  In other words, the intersection of the minimal surfaces $M(\ll)$ and $M(\ll+d\ll)$ happens inside $M_R(\ll)$ and $M_L(\ll+d\ll)$.}}
\end{figure}

The differential entropy associated with a one-parameter family of regions $R(\ll)=[B_L(\ll),B_R(\ll)]$ is defined as:
\be\la{sdiff}
S_\diff= \int \lt\{ S_{EE}[B_L(\ll),B_R(\ll)] - S_{EE}[B_L(\ll+d\ll),B_R(\ll)] \rt\} .
\ee
{Here $S_{EE}[B_L(\ll),B_R(\ll)]$ is the generalized area of the extremal surface $M(\ll)$. When $M(\ll)$ is minimal, \JS{and $[B_L(\ll),B_R(\ll)]$ is a well-defined boundary region, $S_{EE}[B_L(\ll),B_R(\ll)]$} is the entanglement entropy of the ring $R(\ll)$ (in units where $4G_N \equiv 1$).}
Our claim is that formula~(\ref{sdiff}) reconstructs the generalized area \er{garea} of the original surface:
\be\la{claim}
S_\diff = A \,.
\ee

%
%

\se{Proof}
\label{proof}
{The quantity} $S_{EE}[B_L,B_R]$ is computed as the generalized area of the {extremal} surface stretched between $B_L$ and $B_R$, or the on-shell action of
\be
S= \int ds \, d^{d-2}\h \, L(x^\m,\pa_s x^\m,\pa_{\h^\a} x^\m)
\ee
subject to the initial and final conditions $B_L$ and $B_R$.  We may write this {extremal} surface as $M[B_L,B_R]$ and this on-shell action as $S_\on[B_L,B_R]$.
As we vary $\ll$, the minimal surfaces $M(\ll)$ and $M(\ll+d\ll)$ generically intersect on a loop in the bulk.  {Let us call this loop \JS{$\td K(\ll)$};  see Fig.~\re{figproof}. }
By construction we have:
\be\la{obvious}
S_{EE}[B_L(\ll),B_R(\ll)] = S_\on[B_L(\ll),\td K(\ll)] + S_\on[\td K(\ll),B_R(\ll)] \,.
\ee

\begin{figure}[t]
\centering
\includegraphics[width=0.5\textwidth]{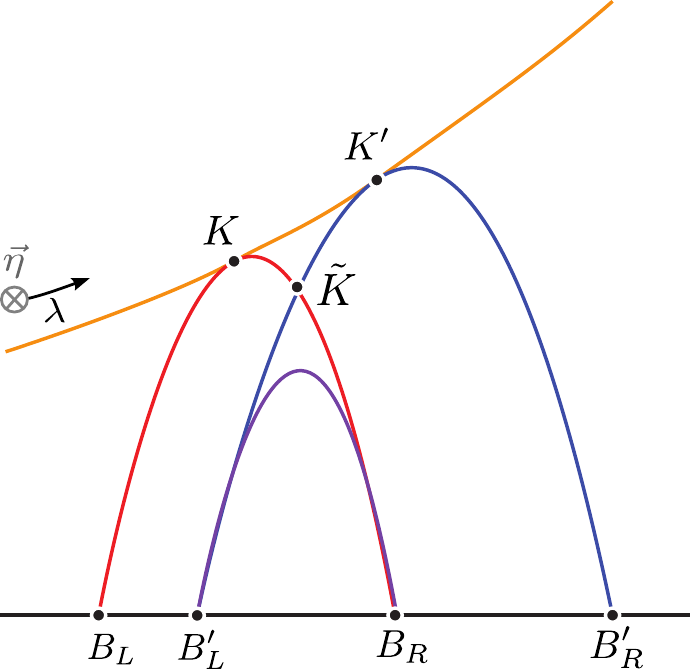}
\caption{\label{figproof}{Nomenclature used in} the proof.  Unprimed symbols denote quantities at $\ll$ {while primes mark} quantities at $\ll+d\ll$.  The other $d-2$ dimensions are denoted by $\vec\h$.}
\end{figure}

First, we claim that
\be\la{replace}
S_{EE}[B_L(\ll+d\ll),B_R(\ll)]= S_\on[B_L(\ll+d\ll),\td K(\ll)] + S_\on[\td K(\ll),B_R(\ll)] \,
\ee
up to terms that are second-order or higher in $d\ll$ and can therefore be neglected.  To prove this, we note that the {extremal} surface $M[B_L(\ll+d\ll),B_R(\ll)]$ stretched between $B_L(\ll+d\ll)$ and $B_R(\ll)$ is different from the union of two {extremal} surfaces $M[B_L(\ll+d\ll),\td K(\ll)]$ and $M[\td K(\ll),B_R(\ll)]$ only by terms that are linear or higher-order in $d\ll$.  After all, these  surfaces come from {extremal} surfaces stretched at the asymptotic boundary whose boundary conditions vary by $\mo(d\ll)$.
It follows that the difference between the left- and right-hand side of \er{replace} vanishes at the linear order in $d\ll$, because the linear term is proportional to the equation of motion, which is satisfied.

We may now plug \er{obvious} and \er{replace} into the differential entropy \er{sdiff}:
\be
S_\diff = \int \lt\{ S_\on[B_L(\ll),\td K(\ll)] - S_\on[B_L(\ll+d\ll),\td K(\ll)] \rt\} \,.
\ee
We may further rewrite the integrand as:
\begin{eqnarray}
S_\on[B_L(\ll),\td K(\ll)] - S_\on[B_L(\ll+d\ll),\td K(\ll)] & = & S_\on[B_L(\ll),K(\ll)] - S_\on[B_L(\ll+d\ll),K(\ll+d\ll)] \nonumber \\
& + & S_\on[K(\ll),\td K(\ll)] + S_\on[\td K(\ll),K(\ll+d\ll)] \,. \la{intg}
\end{eqnarray}
Since the minimal surface $M(\ll)$ at each $\ll$ is tangent to the original surface at $K(\ll)$ by construction, the second line of \er{intg} may be replaced by the generalized area swept out between $K(\ll)$ and $K(\ll+d\ll)$ on the original surface,
\be
S_\on[K(\ll),\td K(\ll)] + S_\on[\td K(\ll),K(\ll+d\ll)] = \lt. \int d^{d-2}\h \, L(x^\m,\pa_\ll x^\m,\pa_{\h^\a} x^\m) \rt|_{K(\ll)} d\ll \,,
\ee
up to {terms second-order or higher} in $d\ll$.  Therefore, \er{intg} becomes:
\bm\la{intg2}
S_\on[B_L(\ll),\td K(\ll)] - S_\on[B_L(\ll+d\ll),\td K(\ll)] = -\fr{d S_\on[B_L(\ll),K(\ll)]}{d\ll}\, d\ll \\ + \lt. \int d^{d-2}\h \, L(x^\m,\pa_\ll x^\m,\pa_{\h^\a} x^\m) \rt|_{K(\ll)} d\ll \,.
\em
{Integrating over $\ll$, we recover the desired result
\be
S_\diff =  \int d\ll \, d^{d-2}\h \, L(x^\m,\pa_\ll x^\m,\pa_{\h^\a} x^\m) = A
\ee
as long as the first term in \er{intg2} does not produce boundary terms. When the 1-parameter foliation of the given surface by $K(\ll)$ is nondegenerate so that $\ll$ is periodic, this is automatic. 
In Sec.~\ref{spheres} we explain the circumstances, under which the same conclusion extends to degenerate foliations, where $\ll$ varies over an interval.}

\se{Examples}
\label{examples}

\begin{figure}[t]
\centering
\includegraphics[width=0.328\textwidth]{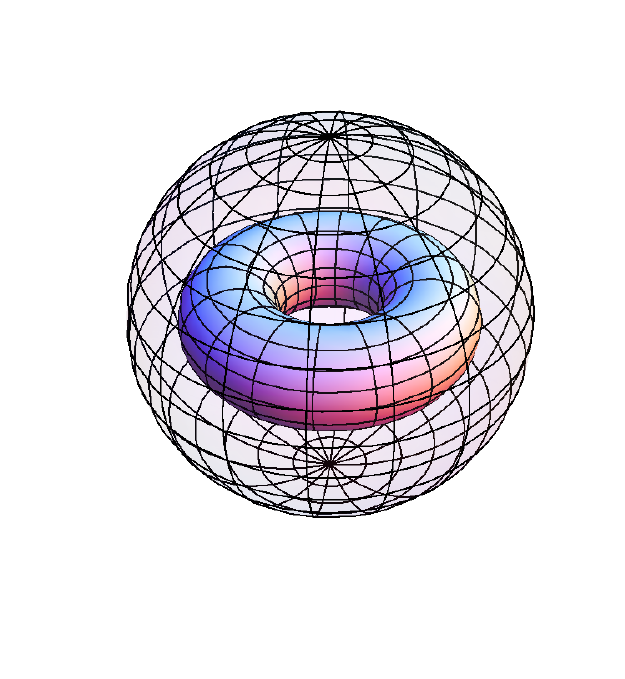}
\includegraphics[width=0.328\textwidth]{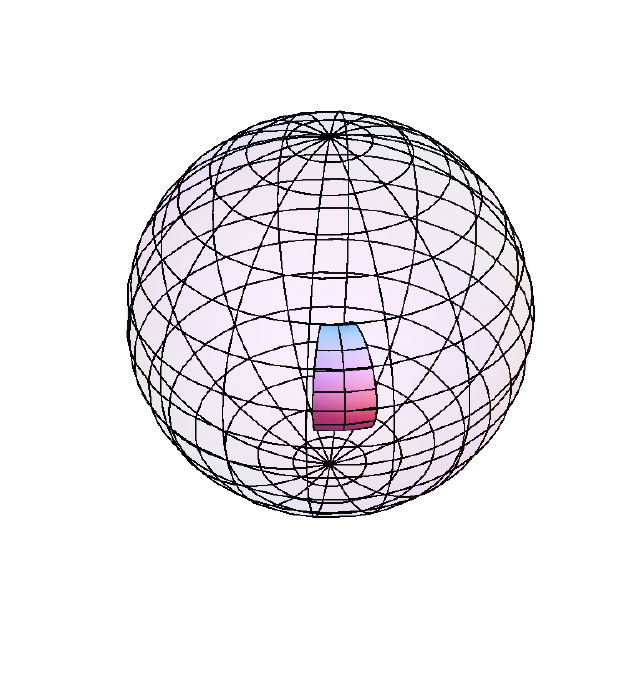}
\includegraphics[width=0.328\textwidth]{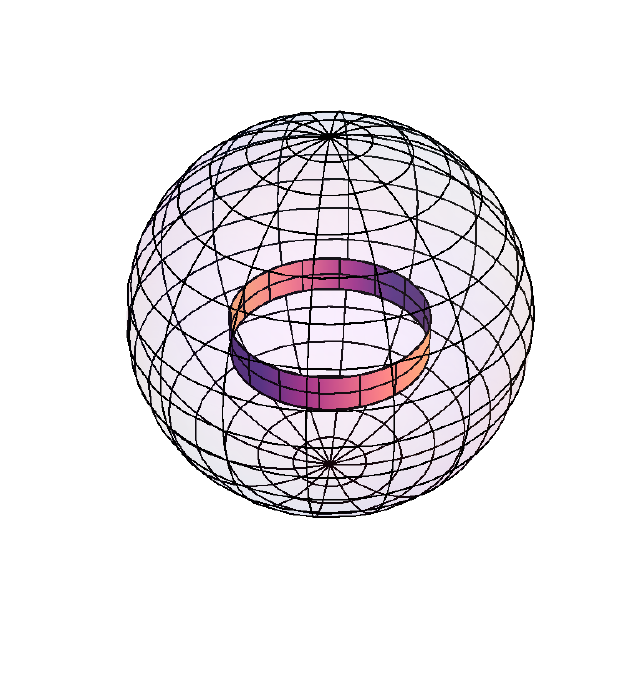}
\caption{\label{figtorus}Left: reconstruction of a torus on a constant time slice of $AdS_4$.  Middle: the infinitesimal version where the loop $K(\ll)$ sweeps out a small area.  Right: we have exploited the symmetry of AdS and moved the loop $K(\ll)$ to the center of AdS.}
\end{figure}

\sse{Torus}
We start with an example where the foliation by $K(\ll)$ is nondegenerate. The simplest instance is a torus on a constant time slice of $AdS_4$; see Fig.~\ref{figtorus}. 


We employ pseudo-spectral methods to numerically solve the partial differential equations that determine the embedding of various {extremal surfaces. Assuming} Einstein gravity, we find the differential entropy \er{sdiff} via the Ryu-Takayanagi formula.  We verify that this gives the area of the torus by numerically showing the infinitesimal form of the claim
\bm\la{inf}
S_{EE}[B_L(\ll),B_R(\ll)] - S_{EE}[B_L(\ll+d\ll),B_R(\ll)] = -\fr{d S_\on[B_L(\ll),K(\ll)]}{d\ll} d\ll \\
+ \lt. \int d^{d-2}\h \, L(x^\m,\pa_\ll x^\m,\pa_{\h^\a} x^\m) \rt|_{K(\ll)} d\ll
\em
as we vary the loop {from} $K(\ll)$ to $K(\ll+d\ll)$.  Using the maximal symmetry of AdS, we may move the loop $K(\ll)$ to a more symmetric point such as the center of AdS simply for the purpose of easing our numerical computation.  A torus with a generic shape and a generic foliation would translate into the statement that the loops $K(\ll)$ and $K(\ll+d\ll)$ are not necessarily perfect circles, of the same size, or parallel to each other.

To give a concrete example, let us focus on a torus that is specified in the following way.  We choose the coordinates of $AdS_4$ such that its metric is
\be
ds^2 = \fr{1}{z^2} \[ -(1+z^2) dt^2 + \fr{dz^2}{1+z^2} + d\q^2 + \sin^2\q d\p^2 \] \,,
\label{ads4metric}
\ee
where we have set the radius of curvature to 1 for simplicity.  Let us start with the circle at $\{z=z_0, \q=\pi/2\}$, where $z_0$ is a constant.  Intuitively, we would like to generate a torus by moving the circle to the ``right'' and rotating it in the orthogonal direction.  This is most easily specified by embedding the constant time slice of $AdS_4$, which is a 3-dimensional hyperbolic space, {in} a 4-dimensional Minkowski space with coordinates $X^\m$, $\m=0,1,2,3$.  Explicitly, the embedding map is:
\be
X^0 = \fr{\sqrt{1+z^2}}{z} \,,\qu
X^1 = \fr{1}{z} \sin\q \cos\p \,,\qu
X^2 = \fr{1}{z} \sin\q \sin\p \,,\qu
X^3 = \fr{1}{z} \cos\q \,.\qu
\ee
We move the circle to the ``right'' by applying a boost in the $(X^0,X^1)$ directions with rapidity $\t$ {and then rotate the circle in the $(X^1,X^3)$ directions. This} gives us a torus in the bulk.  In particular, we get an infinitesimal piece of the torus by rotating the circle by an angle $\d\y$.  The boundary of this infinitesimal piece consists of the two loops we have been calling $K(\ll)$ and $K(\ll+d\ll)$. {The relative magnitude of $\tau$ and $\d\y$ controls the size of the second cycle of the torus.}

As mentioned before, we boost this infinitesimal problem to the center of AdS$_4$ to simplify the computation.  We have numerically solved the infinitesimal case with the following parameters:
\be
z_0 = \fr12 \,,\qqu
\t = 2 \,,\qqu
\d\y = 10^{-3} \,,\qqu
z_c = 10^{-1} \,,
\ee
where $z_c$ is the cutoff near the AdS boundary.  We find numerically that the differential entropy is
\be
S_{EE}[B_L(\ll),B_R(\ll)] - S_{EE}[B_L(\ll+d\ll),B_R(\ll)] \approx 0.1066 \,,
\ee
and the total derivative term is
\be
\fr{d S_\on[B_L(\ll),K(\ll)]}{d\ll} d\ll \approx 0.0486 \,.
\ee
Adding these two contributions, we predict that the area swept out between the two loops $K(\ll)$ and $K(\ll+d\ll)$ is
\be
\d A_{\rm predicted} \approx 0.1552 \,,
\ee
which agrees with the actual area
\be
\d A_{\rm actual} \approx 0.1530
\ee
within numerical errors including those from ignoring higher order terms in $\d\y$.  Details of the numerical computation are collected in Appendix~\re{app}.
{We have numerically verified the infinitesimal claim \er{inf} in tori of other shapes, {which generally have no symmetry}.}

{It should be noted that we may foliate a torus in more than one way. Each foliation gives a different reconstruction of the area of the torus. For example, we may foliate the torus by cycles dual to the $K(\ll)$ used in the discussion above.}

\sse{Sphere}\label{spheres}
In this subsection we consider examples where the bulk surface is topologically a sphere.  
{This case differs from the torus in that any foliation by lower-dimensional ``loops'' necessarily degenerates at two points.}

Let us first consider the example of a perfect $(d-1)$-dimensional sphere inside the $AdS_{d+1}$ bulk.  It is natural to foliate the perfect sphere {by ``loops'' $K(\ll)$, which are perfect $(d-2)$-dimensional spheres.}  One might worry that the extremal surfaces $M(\ll)$ constructed from $K(\ll)$ behave badly by developing caustics as $K(\ll)$ approaches {a degeneration point.}  {However, a caustic is inconsistent with the extremality condition. As $K(\ll)$ approaches a degeneration point, the associated extremal surface $M(\ll)$ develops an increasingly narrow ``neck''.}
In the limit that $K(\ll)$ shrinks to zero size, $M(\ll)$ pinches off at the neck {and splits up into two copies of a special, uniquely determined extremal surface. The latter is selected by two conditions: it is tangent to the bulk sphere at the degeneration point and its extrinsic curvature tensor vanishes identically (not just in the trace).
The variation of the surfaces $M(\ll)$ with $\ll$ is illustrated in Fig.~\re{figsphere}.}

\begin{figure}[t]
\centering
\includegraphics[width=0.328\textwidth]{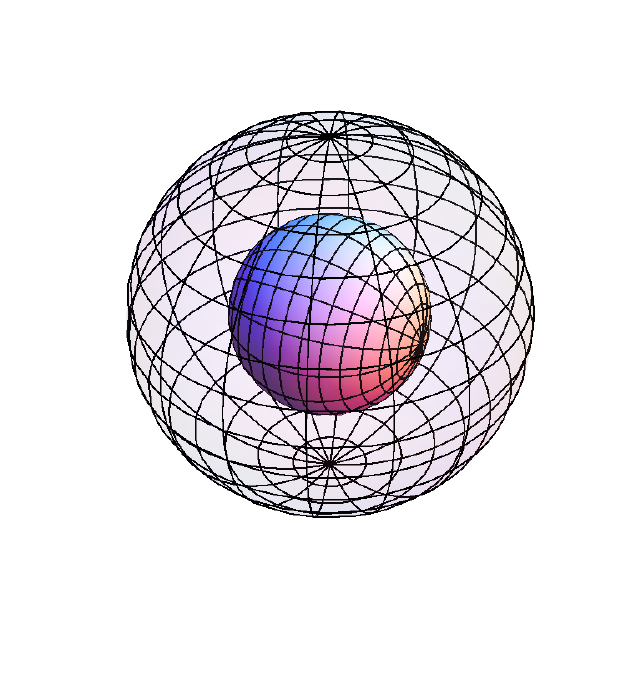}
\includegraphics[width=0.328\textwidth]{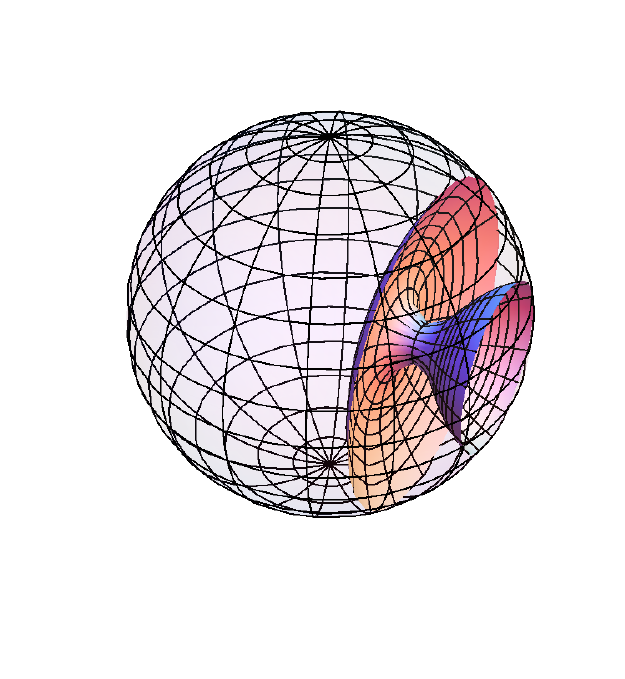}
\includegraphics[width=0.328\textwidth]{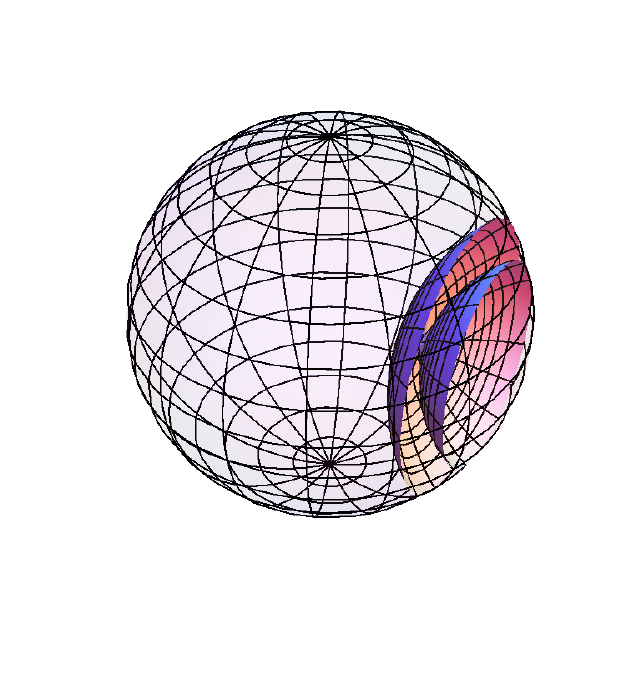}
\caption{\label{figsphere}Left: foliation of a sphere by circles that degenerate at two points.  Middle and right: the extremal surfaces $M(\ll)$ constructed from circles $K(\ll)$ as they approach a degeneration point.}
\end{figure}

The infinitesimal form \er{intg2} of our claim naturally holds in this case. {We may therefore use the differential entropy to reconstruct the generalized area of the sphere. This conclusion is subject to two caveats discussed below.}

First, due to the degeneration points the $\ll$ direction is no longer closed, so we pick up a boundary term at each degeneration point:
\be
S_\diff = A - S_\on[B_L(\ll),K(\ll)] \Big|^{\ll=\ll_R}_{\ll=\ll_L} \,,
\ee
where $\ll_L$ and $\ll_R$ are the coordinates of the two degeneration points.  According to our earlier discussion, the boundary term $S_\on[B_L(\ll),K(\ll)]$ at a degeneration point is simply the on-shell action of the unique extremal surface that is tangent to the sphere at the degeneration point and {has a component-wise vanishing extrinsic curvature tensor}.  For a perfect sphere it is obvious that the boundary terms cancel between the two degeneration points (as long as the UV cutoff of the theory also respects the spherical symmetry). {We therefore obtain the desired result:}
\be\la{claim2}
S_\diff = A \,.
\ee

More generally, we may go beyond perfect spheres and consider any surface that is topologically a sphere.  If the surface admits a $\mathbb{Z}_2$ symmetry, we can choose a foliation that respects the $\mathbb{Z}_2$ symmetry in the sense that any loop $K(\ll)$ is mapped to some other loop $K(\ll')$ under the $\mathbb{Z}_2$.  Then the two degeneration points must be mapped to each other under the $\mathbb{Z}_2$ symmetry, {so their boundary terms cancel (as long as the UV cutoff of the theory is $\mathbb{Z}_2$ symmetric) and eq.~\er{claim2} follows.}  {Note that the same argument applies to $\mathbb{Z}_2$-symmetric surfaces of arbitrary topology: if we choose a foliation and UV cutoff that respect the $\mathbb{Z}_2$ symmetry, and none of the degeneration points is $\mathbb{Z}_2$-invariant (so that they are mapped pairwise by $\mathbb{Z}_2$), the boundary terms arising at the degeneration points necessarily cancel and we have eq.~\er{claim2} for these surfaces including higher-genus Riemann surfaces and their higher-dimensional generalizations.}

Without any symmetry, we may still {locate the degeneration points at two distinct points on the surface with canceling boundary terms.  Indeed,} the boundary term $S_\bdy\eq S_\on[B_L(\ll),K(\ll)]$ at a degeneration point is completely determined by the location and tangent plane of that point, independent of the details of the foliation elsewhere.  Therefore $S_\bdy$ is a continuous function on a closed surface {and it is always possible to find two distinct points with equal values of $S_\bdy$.  We may then construct a foliation, which degenerates at exactly these two points. This construction ensures that eq.~\er{claim2} holds.  It should be noted, however, that the determination of $S_\bdy$ and the selection of two points with equal $S_\bdy$ is sensitive to the choice of the geometric} cutoff surface near the asymptotic boundary of the bulk geometry.

The second caveat is that the extremal surfaces $M(\ll)$ {cease to be} minimal as the circle $K(\ll)$ comes close to a degeneration point. We discuss this in detail in the next Section.

\se{Extremal but not minimal surfaces}
\label{extr}

As illustrated in Fig.~\re{fignonmin}, in AdS$_4$ there are up to three extremal surfaces that asymptote to the same ring-shaped region. Of course, only one of them -- that with the smallest area -- computes the entanglement entropy of a spatial region in the boundary field theory. The boundary interpretation of the others is not currently known. Ref.~\cite{Czech:progress} addresses this question in the context of conical defect and BTZ spacetimes in three bulk dimensions, where extremal but nonminimal curves first arise. In this section, we discuss briefly the ``phase structure'' of extremal surfaces in AdS$_4$, the circumstances under which extremal nonminimal surfaces are necessary to describe a given bulk surface, as well as generalizations beyond AdS$_4$. We will also comment briefly on possible boundary interpretations of extremal surfaces. For a fuller discussion consult \cite{Czech:progress}. 

\begin{figure}[t]
\centering
\includegraphics[width=0.328\textwidth]{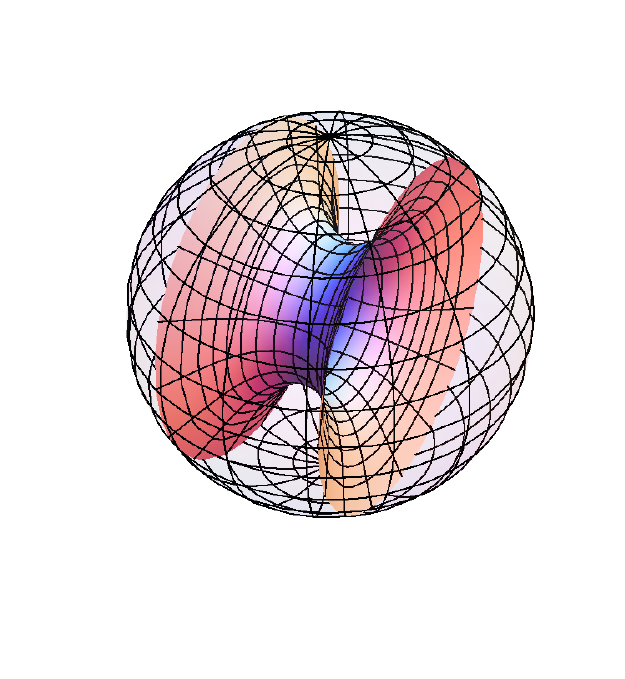}
\includegraphics[width=0.328\textwidth]{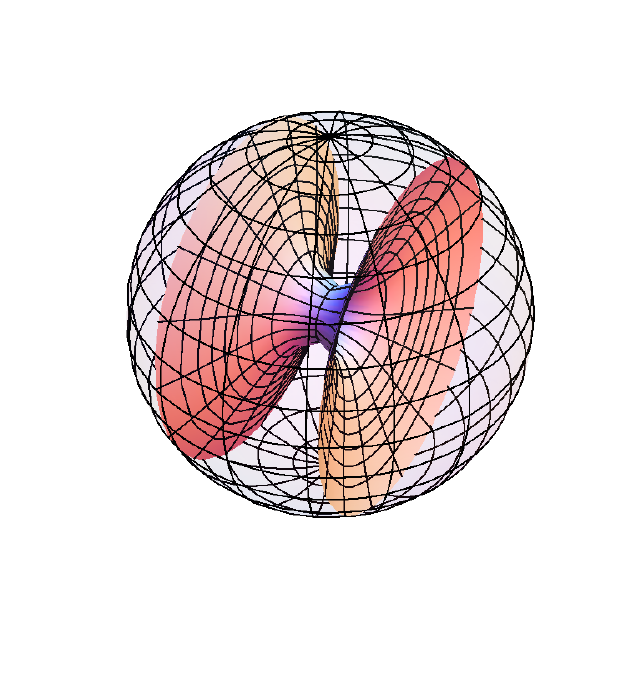}
\includegraphics[width=0.328\textwidth]{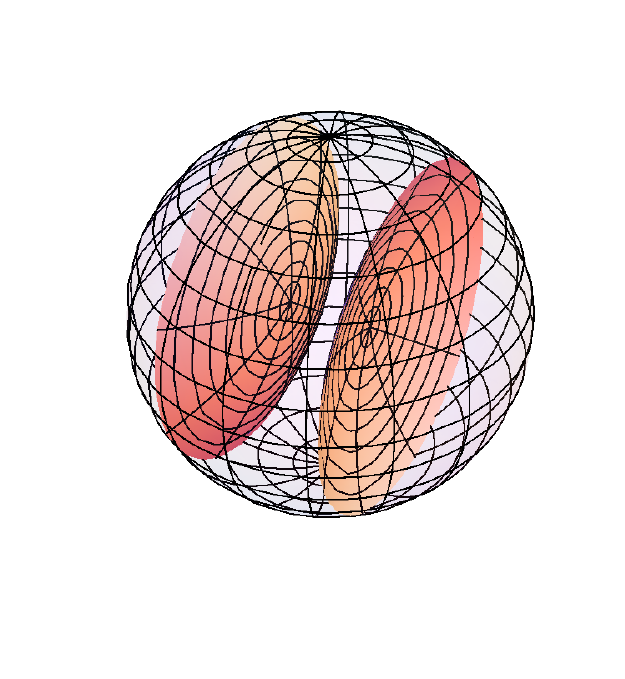}
\caption{\label{fignonmin}Three possible extremal surfaces that asymptote to the same ring-shaped region on the asymptotic boundary: thick, thin, and disconnected.}
\end{figure}

\subsection{Phase structure of extremal surfaces}
\label{secphases}
\begin{figure}[t]
\centering
\includegraphics[width=0.496\textwidth]{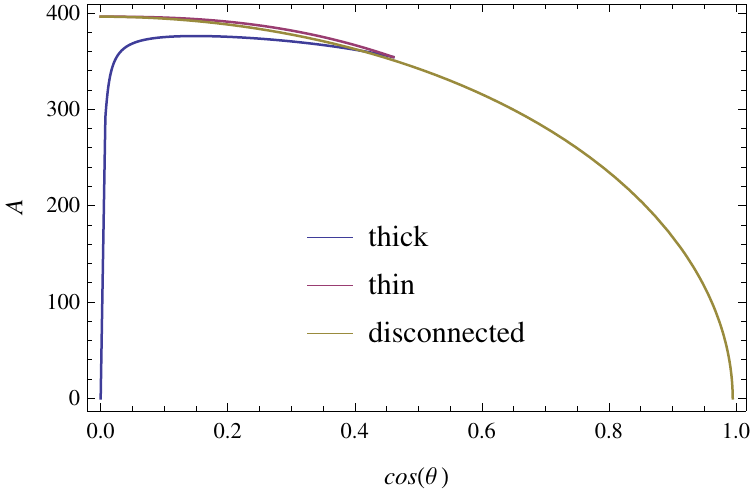}
\includegraphics[width=0.496\textwidth]{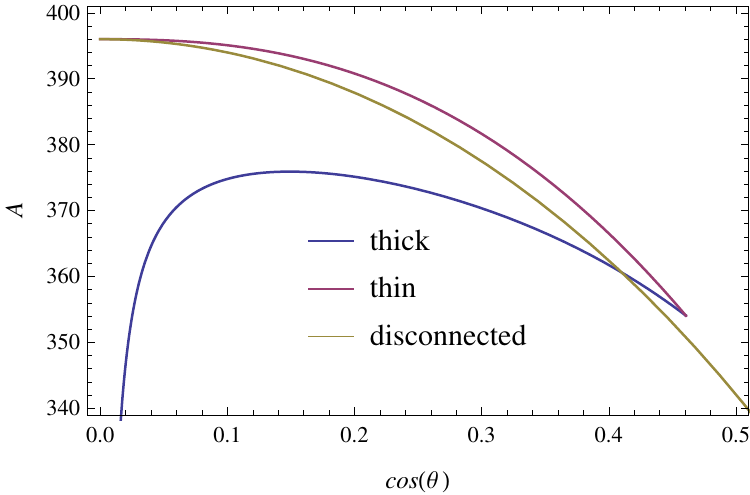}
\caption{\label{figphase}Plots of the areas of various extremal surfaces in AdS$_4$ as functions of the width of the boundary region.  The right plot is a magnification of the upper-left corner of the left plot.  To show the different curves clearly, we have slightly exaggerated their differences without changing the qualitative features of the plots.  The cutoff is set to $z_c=5\times10^{-3}$.}
\end{figure}

Fig.~\re{fignonmin} displays the extremal surfaces in AdS$_4$, which asymptote to longitudinal circles located at latitudes $\theta$ and $\pi-\theta$ on the asymptotic boundary (see metric~\ref{ads4metric}). Starting from $\theta = \pi/2$ (the equatorial plane), there are three extremal surfaces, two of which are connected. According to the \JS{diameter} of the connecting \JS{cylindrical region}, we call the connected surfaces ``thick'' or ``thin''. The area of the thin surface is always larger than the area of the thick surface resting on the same boundary region. As we vary $\theta$ away from $\pi/2$, the thick and thin surfaces approach the same limiting shape with \JS{diameter} of order $L_{\rm AdS}$ ($z_c = 0.700$), which is reached at the critical value $\theta \approx \cos^{-1} 0.46$. Beyond that (for $\theta < \cos^{-1} 0.46$) there is only one extremal surface, which consists of two identical, disconnected pieces. This type of surface exists for all values of $\theta$, but it is minimal only for $\theta < \cos^{-1} 0.41$, where a first order transition occurs. For $\cos^{-1} 0.41 < \q < \pi/2$ the area of the disconnected surface lies between the thick and thin connected areas, equaling the latter at $\pi/2$, where the thin connected surface pinches off into two disconnected components. The full phase structure of extremal surfaces in AdS$_4$ is summarized in the diagram in Fig.~\ref{figphase}.

When the boundary ``ring'' does not lie between two identical circles but assumes a generic shape, the extremal surfaces depend on all the parameters of the boundary shape. In this case, drawing a complete phase diagram is not possible. Nevertheless, the phase structure is qualitatively the same as in Figs.~\ref{fignonmin}-\ref{figphase}.


\subsection{Relevance of extremal but nonminimal surfaces}

Consider a bulk surface foliated by loops $K(\ll)$, which degenerate at {finitely  many points}. We discussed the simplest example, a sphere, in Sec.~\ref{spheres}.  If the characteristic size of the surface is comparable or larger than the AdS curvature scale, we can assume that a generic loop $K(\ll)$ selects a tangent extremal surface, which is of the thick type. As seen on Fig.~\ref{figphase}, the thick extremal surface is generically minimal, so it computes the entanglement entropy of a spatial region on the boundary. Thus, the integrand in eq.~(\ref{sdiff}) at generic $\ll$ involves entanglement entropies of spatial regions in the boundary field theory.

This conclusion does not hold in a neighborhood of a degeneration point of the foliation. In the limit that $\ll$ approaches a degeneration point, the loops $K(\ll)$ shrink down to a point. Following this shrinkage, the extremal surfaces that are tangent to the given bulk surface at $K(\ll)$ cannot remain in the thick regime. Instead, in a neighborhood of a degeneration point the tangent extremal curves are of the thin, nonminimal type. At the degeneration point itself, the tangent surface is the limiting surface on which the thin and disconnected branches meet up (\XD{$\theta = \pi/2$} in Fig.~\ref{figphase}). In this way, a degeneration of the foliation necessarily involves nonminimal surfaces in the differential entropy formula. Although the same bulk curve can be captured by formula~(\ref{sdiff}) in many ways, the degeneration of a foliation is mandated by topology, so it cannot be evaded by a clever redefinition of the foliation.

In the preceding paragraphs, we assumed that a generic loop $K(\ll)$ is no smaller than the AdS curvature scale. Another context in which nonminimal extremal surfaces make an appearance is when we attempt to localize a bulk surface in a region much smaller than $L_{\rm AdS}$. Such small regions are not covered by thick extremal surfaces, so thin extremal surfaces again become necessary. This case is distinct from degenerate foliations, as exemplified by a small bulk torus whose both fundamental cycles are much smaller than the curvature scale of the ambient AdS space.

\begin{figure}[t]
\centering
\vspace{-1cm}
\includegraphics[width=0.496\textwidth]{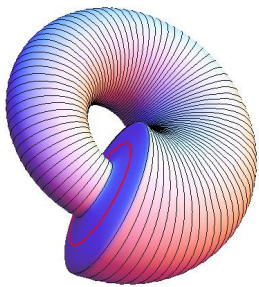}
\vspace{-.7cm}
\caption{\label{funnytorus}A surface, which is large and has a nondegenerate foliation, whose
differential entropy description nevertheless involves a nonminimal
extremal surface. The leaf of the foliation highlighted in red is
tangent to an extremal surface in the disconnected phase, which (for a
generic choice of parameters) is nonminimal.}
\end{figure}

One may imagine more general scenarios, in which extremal but not minimal surfaces enter eq.~(\ref{sdiff}). \BC{For example, Fig.~\ref{funnytorus} shows a topological torus with a nondegenerate foliation, which is everywhere large compared to $L_{\rm AdS}$ and nevertheless becomes tangent at a loop $K(\ll)$ to a nonminimal extremal surface.} While it may be possible to eliminate extremal but not minimal tangent surfaces from formula (\ref{sdiff}) by redefining the foliation, degenerate foliations and surfaces confined to a small bulk region are two cases, where this is definitely not possible.

To avoid dealing with nonminimal surfaces, it is possible to start from the field theory side and {\it construct} bulk surfaces holographically. To do so, consider a family of field theory regions parameterized by $\lambda$, each of which has two disconnected boundaries $B_{L,R}(\lambda)$. Their ``outer envelope'' \cite{Myers:2014jia} is a bulk surface, which may fail to be differentiable in two ways. First, the extremal surfaces $M(\ll)$ may not vary continuously with $\ll$; this happens when there is a first order phase transition in entanglement entropy (see \cite{renyis, plateaux} for examples). Second, the ``outer envelope'' may develop caustics (cusps); for examples see \cite{Veronika, Wienthesis, Headrick:progress, lampros}. At caustics, the differential entropy may undergo a sign change, which can be understood as trading the boundary regions parameterized by $\ll$ for their complements \cite{Headrick:progress, lampros}. The problem of reconstructing the bulk surface starting from a family of boundary regions has been discussed in \cite{Wienthesis, Headrick:progress}. Note that such a construction has an added bonus: we need not assume that the Cauchy problem discussed in Sec.~\ref{strategy} has a solution for all $\lambda$.

\subsection{Boundary interpretation}
\label{extrdisc}
A common lore in AdS/CFT is that it is difficult to achieve bulk locality on scales smaller than the AdS curvature radius. To holographically zoom in on a sub-AdS region, one must study a
sector of the field theory, which contains no data about the spatial
organization of its degrees of freedom, namely a matrix quantum
mechanics. How to reconcile this with the Ryu-Takayanagi proposal and
formula (\ref{sdiff}), which relates surface areas to entanglement entropies
-- quantities, which were invented to quantify correlations between
spatially separated degrees of freedom?


We anticipate that extremal but nonminimal surfaces bridge the gap between these two points of view. The success of formula (\ref{sdiff}) in reproducing sub-AdS scale areas using nonminimal surfaces provides partial evidence for this qualitative statement. Another piece of evidence comes from considering RG flows: the distinction between a minimal and a merely extremal surface can depend on the large scale cutoff in gravity. We infer from this that the field theory contains quantities, which (a)  are computed in holography by areas of extremal yet nonminimal surfaces and (b) reduce to entanglement entropies when the field theory flows to the deep infrared. These putative quantities should characterize how internal (matrix) degrees of freedom depend on large scales in field theory, reducing to a complete description of the matrix quantum mechanics.

For a more complete discussion of nonminimal extremal surfaces, we refer the reader to \cite{Czech:progress}. Set in the context of AdS$_3$/CFT$_2$, it gives a definition of a novel boundary quantity called ``entwinement,'' whose bulk dual is the area of an extremal (not necessarily minimal) surface. Qualitatively, it captures (an analogue of) the entanglement between gauge equivalent subsectors of the field theory, compounded over the entire asymptotic boundary.
The definition of entwinement relies on special properties of 1+1-dimensional conformal field theories, so it is not clear how to lift it to higher dimensions. The results of the present paper motivate the problem of extending entwinement to higher-dimensional holographic theories. 

\se{Discussion}
\label{discussion}
The Ryu-Takayanagi proposal \cite{Ryu:2006bv} brought to light a new way of thinking about the emergence of the gravitational spacetime in holography, which treats boundary entanglement as the basic ingredient. The differential entropy formula
\be
S_\diff= \int \lt\{ S_{EE}[B_L(\ll),B_R(\ll)] - S_{EE}[B_L(\ll+d\ll),B_R(\ll)] \rt\} = A / 4G
\label{mainagain}
\ee
combines these ingredients (and their appropriate generalizations, viz. Sec.~\ref{extr}) to reproduce the area of an arbitrary surface in the bulk. It was originally written for pure AdS$_3$ \cite{Balasubramanian:2013lsa} and later extended to a variety of holographic contexts \cite{Myers:2014jia}. This paper, which draws on a proof of eq.~(\ref{mainagain}) given in \cite{Headrick:progress}, shows that the relation between differential entropy and areas of arbitrary surfaces extends to all bulk curves, which admit a 1-parameter foliation over a closed manifold. Subtleties associated with possible degeneration points of the foliation were discussed in Sec.~\ref{spheres}.

Before closing, we highlight some aspects of  formula~(\ref{mainagain}), which have become apparent in the present, higher-dimensional context.

\begin{figure}[t]
\centering
\includegraphics[width=0.45\textwidth]{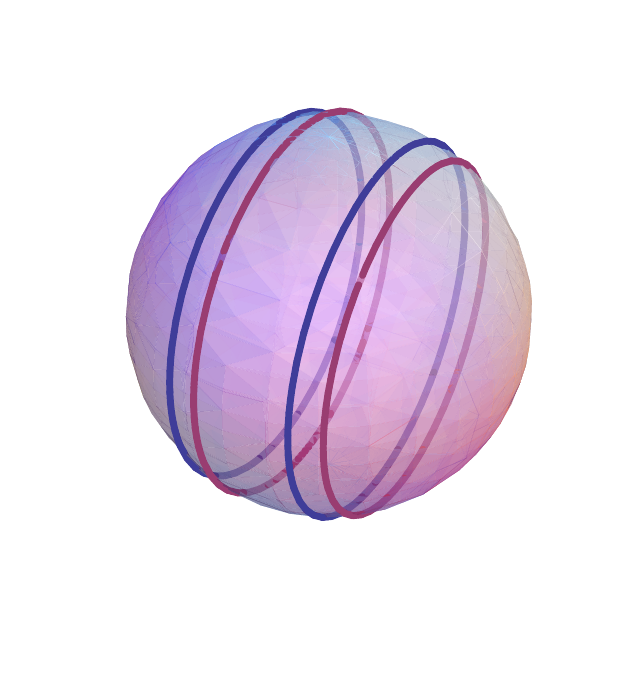}
\includegraphics[width=0.45\textwidth]{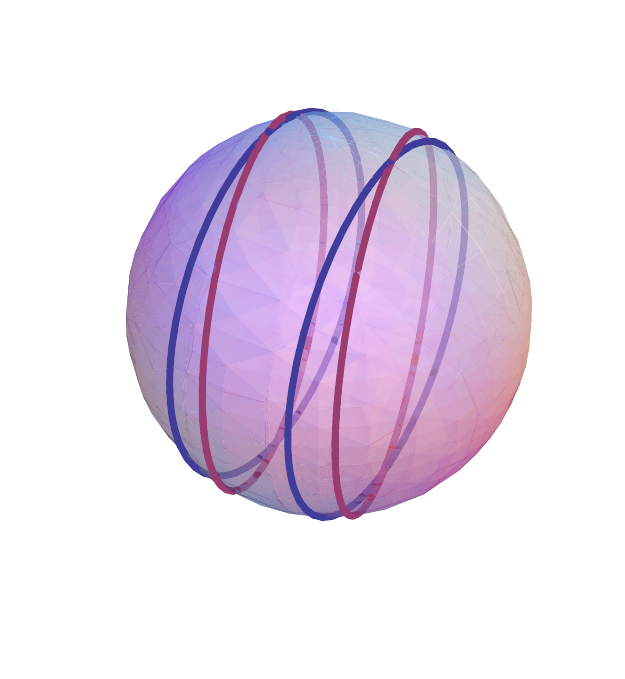}
\caption{\label{figring}The region $[B_L(\ll+d\ll),B_R(\ll)]$ is same as the intersection of $R(\ll)$ and $R(\ll+d\ll)$ on the left, but they are different on the right.  Here the blue loops are $B_{L,R}(\ll)$ and the purple ones are $B_{L,R}(\ll+d\ll)$.}
\end{figure}

\paragraph{Beyond unions and intersections of boundary regions}
Assuming that the subtlety discussed in Sec.~\ref{spheres} does not arise, the first term in the integral (\ref{mainagain}) computes the entanglement entropy of a ring-like region $R_\lambda$, which lies between two boundary loops $B_L(\lambda)$ and $B_R(\ll)$. In AdS$_3$/CFT$_2$, we can write the second term as the entanglement entropy of the intersection (resp. union) of $R_\lambda$ and $R_{\ll + d\ll}$, depending on the sign of $dB_{L,R}/d\ll$ \cite{Myers:2014jia}:
\be
S_\diff = \int \lt( S_{EE}(R_\ll) 
- \lt\{ \begin{array}{lr} S_{EE}(R_\ll \cap R_{\ll + d\ll}) & ~{\rm if}~B'_L(\ll), B'_R(\ll) > 0 \\
S_{EE}(R_\ll \cup R_{\ll + d\ll}) & ~{\rm if}~B'_L(\ll), B'_R(\ll) < 0 \end{array} \rt. \rt)
\ee
Here we point out that in higher dimensions a rewriting of this type may not in general be possible. It applies only when $B_L(\ll+d\ll)$ and $B_R(\ll)$ do not intersect. In more than three bulk dimensions, it is easy to envision an example where these two loci do in fact intersect, see Fig.~\ref{figring}. When this occurs, the intersection of the surfaces $S_{EE}[B_L(\ll), B_R(\ll)]$ and $S_{EE}[B_L(\ll+d\ll), B_R(\ll+d\ll)]$, which we dubbed $\tilde{K}(\ll)$, extends all the way to the boundary. Because $\tilde{K}(\ll)$ for infinitesimal $d\ll$ defines the bulk surface, this is the case of a generic foliation of a surface that reaches all the way to the boundary. 

\paragraph{Degenerate foliations}
{Eq.~(\ref{mainagain}) depends on the foliation $K(\ll)$ of the bulk surface. If the foliation is degenerate, as in the case of a sphere discussed in Sec.~\ref{spheres}, an additional subtlety arises. In this situation the equality~(\ref{mainagain}) applies only up to boundary terms evaluated at degeneration points.} {Where symmetry cannot be used to guarantee the vanishing of boundary terms at the degeneration points, this must be added as an additional constraint on the choice of foliation.} \JS{Our construction can accommodate any such foliation that degenerates at only finitely many $\ll$. A foliation with infinitely many degenerate leaves does not necessarily locally generate the area of the surface as in Eq. \er{intg2}.}

\paragraph{Integrand need not be finite}
The canceling of boundary terms at degeneration points is sensitive to the choice of UV cutoff. This fact illustrates another general lesson about differential entropy: that the integrand in eq.~(\ref{sdiff}) need not be finite. It is equal to the area element on the bulk surface only up to a total derivative term, which is generically divergent. For example, at a degeneration point it equals the divergent area of a certain tangent extremal surface. Even if the integrand of eq.~(\ref{sdiff}) is made finite in one regularization scheme, it will diverge in different regularization schemes.

\paragraph{Nonminimal extremal surfaces}
In Sec.~\ref{extr} we discussed at length the appearance of extremal but nonminimal surfaces in differential entropy. That point carries a salient lesson: entanglement entropies of spatial regions in the boundary theory are not enough to reconstruct the bulk spacetime. Here we add to this discussion another, related caveat: formula~(\ref{sdiff}) applies only when the tangent extremal surfaces $M(\ll)$ are of the form $M[B_L(\lambda), B_R(\lambda)]$, so that left and right boundary conditions in the field theory can be independently varied. In a spacetime with horizons, the horizon also represents a boundary on which an extremal surface may end. Thus, it may happen that as we follow the foliation parameter $\ll$, the requisite boundary condition $B_L(\ll)$ for tangent surfaces ``jumps'' from the asymptotic boundary onto the horizon. The simplest example of this occurs for nearly radial curves in the BTZ geometry \cite{lampros}, whose differential entropy description involves geodesics with one end on the asymptotic boundary and the other on the horizon. Such geodesics are known to compute two-sided correlators in the thermofield double description of the thermal state \cite{Kraus:2002iv}. The relevance of two-sided quantities to reconstructing bulk objects in a single asymptotic region is puzzling. We hope to return to this problem in the future.

%

\paragraph{A coordinate-independent differential entropy?}
Arguably the least satisfactory aspect of formula~(\ref{sdiff}) is that it singles out a specific foliation of the bulk surface. Said differently, the integral in (\ref{sdiff}) sweeps over only one variable of integration, treating the entanglement entropies $S_{EE}[B_L(\lambda), B_R(\lambda)]$ and their corresponding minimal surfaces as given. It would be more satisfying to covariantize (\ref{sdiff}), so that the integral is done over $d-1$ parameters that stand on an equal footing. One could try to iterate our construction and rewrite $S_{EE}[B_L(\lambda), B_R(\lambda)]$ itself as an integral over a parameter $\lambda_2$, but this approach does not lead to simplifications.

At the same time, we believe that a covariant generalization of formula (\ref{sdiff}) exists. One hint is that for a given bulk surface, any 1-parameter foliation satisfying our (relatively lax) assumptions gives an independent way to recover the area of the surface from boundary data. This means that in more than three bulk dimensions we have an enormous freedom of how to represent a bulk surface in the boundary theory. It would be surprising if this freedom could not be exploited to obtain a covariant prescription.

\section*{Acknowledgements} We thank Vijay Balasubramanian, Jan de Boer, Matthew Headrick, Nima Lashkari, and Robert Myers for valuable discussions.  We also thank Jorge Santos for his invaluable help with numerical computations and the JHEP reviewer for clarifying the point in footnote 1.  This work was supported by the National Science Foundation under grant PHY-0756174.

\appendix
\se{Numerics}
\la{app}

Here we briefly describe the methods and results of numerical computation outlined in Sec.~\re{examples}.  In particular, we will see numerical stability.

We employ pseudo-spectral methods to solve the partial differential equation governing the embedding of the minimal surface, $z(y\eq\cos\q,\p)$, subject to Dirichlet boundary conditions.  We use the Fourier basis in the $\p$ direction and keep the lowest $n_\p=32$ modes.  We choose the Chebyshev basis in the $y$ direction and allow the number of kept modes $n_y$ to vary.

For each $n_y$ we may solve the partial differential equation and calculate the differential entropy.  The predicted area $\d A_{\rm predicted}$ approaches the actual area $\d A_{\rm actual}$ as we increase $n_y$ within the window $32 \le n_y \le 192$, as may be seen from Fig.~\re{fignum}.

To see numerical stability, we plot $\d z_{\max} / \d n_y$ as a function of $n_y$, where $z_{\max}$ denotes the maximum of the solution $z(y,\p)$ for a given $n_y$.  We expect that for stable numerics, $\d z_{\max} / \d n_y$ should decrease exponentially as a function of $n_y$.  This is indeed true, as may be seen from Fig.~\re{figerr}.

\begin{figure}[t]
\centering
\includegraphics[width=0.5\textwidth]{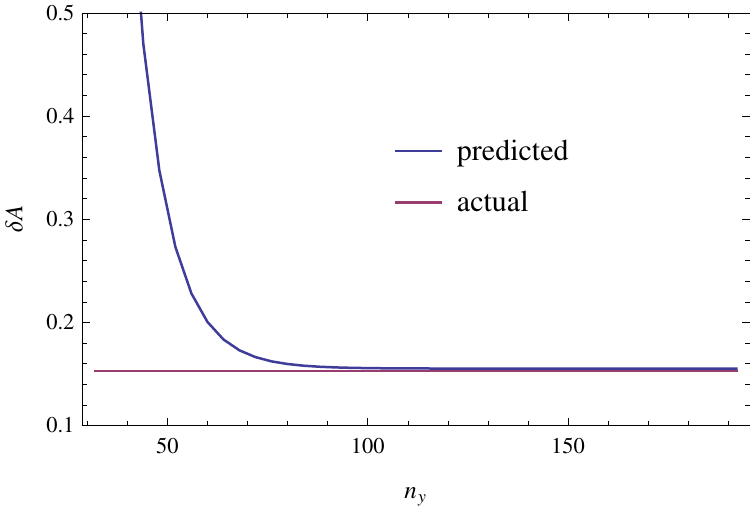}
\caption{\label{fignum}Plot of the predicted area versus the number of kept modes in the $y$ direction.  The horizontal line denotes the actual area.}
\end{figure}
\begin{figure}[h!]
\centering
\includegraphics[width=0.5\textwidth]{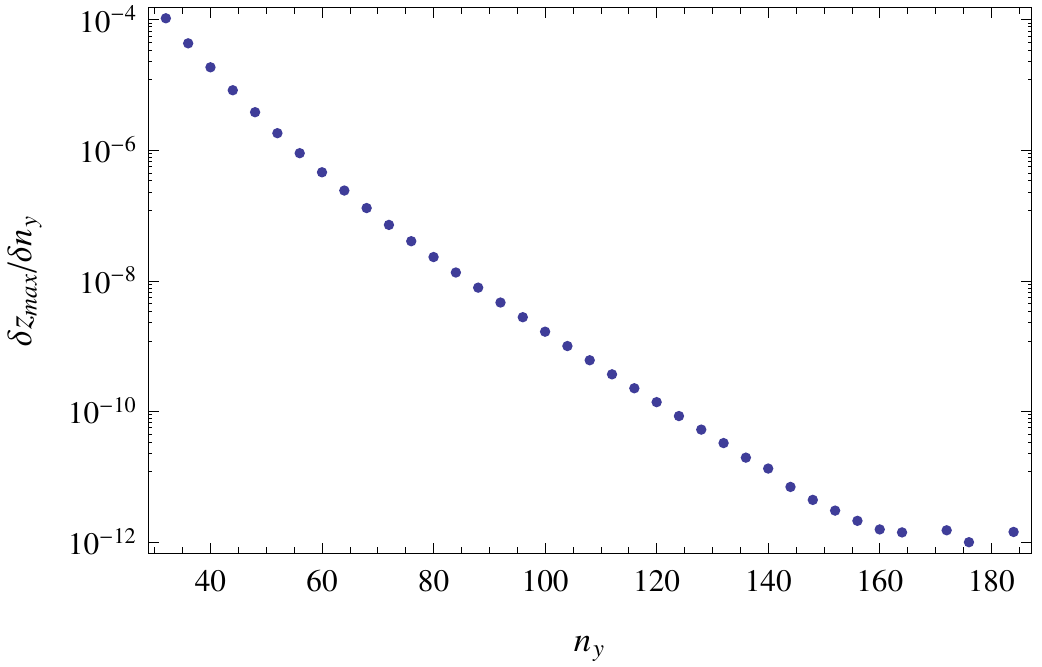}
\caption{\label{figerr}Log plot of $\d z_{\max} / \d n_y$ versus the number of kept modes in the $y$ direction.  The approximately exponential behavior shows numerical stability.}
\end{figure}


\bibliographystyle{JHEP}
\bibliography{references}

\end{document}